\documentclass{aa}  
\usepackage{graphicx}
\usepackage{natbib}
\usepackage{multirow}

\begin{document}

   \title{Structure and variability in the corona of the ultrafast rotator LO~Peg }
   \author{S. Lalitha$^{1}$, J. H. M. M. Schmitt$^{2}$  \and K. P. Singh$^{3}$}
   \authorrunning{S. Lalitha et al.}
   \titlerunning{LO Peg - outer atmospheres}
   \offprints{lalitha.sairam@iiap.res.in}

   \institute{$^{1}$Indian Institute of Astrophysics, Koramangala, Bangalore 560034, India\\
$^{2}$Hamburger Sternwarte, University of Hamburg,
              Gojenbergsweg 112, 21029 Hamburg, Germany\\
              $^{3}$Tata Institute of Fundamental Research, Homi Bhabha road, Mumbai 400005, India \\
              \email{lalitha.sairam@iiap.res.in}
               }
   \date{Received XXXX; accepted XXXX}

 
\abstract
{Low-mass ultrafast rotators show the typical signatures of magnetic activity
and are known to produce flares, probably as a result of magnetic reconnection.  As a 
consequence, the coronae of these stars exhibit very large 
X-ray luminosities and high plasma temperatures, as well as a pronounced inverse FIP effect.}
{To probe the relationship between the coronal properties with spectral type of 
ultra-fast rotators with P$_{rot}< 1d$, we analyse the K3 rapid-rotator LO Peg in 
comparison  with other low-mass rapid rotators of spectral types G9-M1.}
{We report the results of a $42$~ks long \emph{XMM-Newton} observation of LO~Peg and
investigate the temporal evolution of coronal properties like the temperatures, emission measures, 
abundances, densities and the morphology of the involved coronal structures. In addition, we also use the 
\emph{XMM-Newton} data from a sample of rapid rotators  and compare their 
coronal properties to those of LO~Peg.}
{We find two distinguishable levels of activity in the XMM-Newton observation of LO~Peg, which 
shows significant X-ray variability both in phase and amplitude, implying the presence 
of an evolving active region on the surface. 
The X-ray flux varies by $\sim28\%$, possibly due to rotational modulation. 
 During our observation a large X-ray flare
with a peak X-ray luminosity of $\sim2\times10^{30}$ erg/s 
and a total soft X-ray energy release of 7.3$\times10^{33}$ erg was observed.  
Further, at the onset of the flare we obtain clear signatures for the
occurrence of the Neupert effect. 
During the flare a significant emission measure increase in the hotter plasma component is observed, 
while the emission measure in the cooler plasma component is only marginally affected, indicating that 
different coronal structures are involved. 
The flare plasma also shows an enhancement of iron by a factor of $\approx$~2 during the rise 
and peak phase of the flare.
The electron densities 
measured using the \ion{O}{vii} and \ion{Ne}{ix} triplets 
during the quiescent and flaring state are $\approx$ 6~$\times10^{10}$ cm$^{-3}$ 
and  9$\times10^{11}$ cm$^{-3}$, respectively, 
and the large errors prevent us from finding significant density differences 
between quiescent and flaring states. 
Our modeling analysis suggests 
that the scale size of the flaring X-ray plasma is smaller than 0.5~R$_{\star}$. Further, 
the flare loop length appears to be smaller than the pressure scale height of the flaring plasma.  
Our studies show that the X-ray properties of the LO~Peg are very similar to those
of other low-mass ultrafast rotators, i.e., the X-ray luminosity is very close
to saturation, its coronal abundances follow a trend of increasing abundance 
with increasing first ionisation potential, the so-called inverse FIP effect. 
}

\maketitle
%

\section{Introduction}

The Sun is usually considered as a prototype of a low mass star and we often 
extrapolate our knowledge of the properties of the Sun to interpret 
observations of other stars. 
Given that our Sun, a middle-aged main-sequence star, has an X-ray emitting 
corona, the question arises whether the solar corona serves as
a good proxy also for other stars.
Indeed, solar-like stars as a class have been found to be X-rays emitters 
with X-ray luminosities in the range of 10$^{26-31}$~erg/s \citep{rosner_1985}. 
Furthermore, a few low mass stars are considerably more X-ray luminous than the Sun,
and these so-called active low-mass stars are very often also rapid rotators 
\citep{pallavicini_1981} with saturated coronae, i.e., 
the X-ray luminosities of these stars scale with their bolometric
luminosities such that $L_X \approx 10^{-3} L_{bol}$. 
As a result, the X-ray luminosities of active stars of spectral type K and M type range 
from 10$^{28}$ to 10$^{30}$ erg/s \citep{rosner_1985},
raising the question in what respect the coronae of these stars differ 
from the solar corona and how they accommodate the observed excess X-ray emission. 
The X-ray emission from a stellar corona is believed to come from an optically 
thin plasma and therefore scales with the (volume) emission measure $ EM = n^2 V $, 
where $n$ denotes density and $V$ the coronal volume.  
Stars with larger X-ray luminosities than the Sun therefore must have
coronae with a larger volume or a larger coronal density or possibly both.

Active stars also possess magnetic fields which manifest themselves 
in observable features like star spots, flares, emissions in activity-sensitive 
lines like those from Ca~II H\&K, H$\alpha$, etc., large luminosities 
in the X-ray and EUV regime and, last but not least, activity cycles 
\citep{pizzolato_2003}. 
Two well-known examples of such extremely active, ultra-fast rotators are  
BO~Mic and AB~Dor, both of which have been previously studied at X-ray wavelengths in great detail (cf., \citealt{wolter_2008, lalitha_2013}) and both of which
show moderate flares at X-ray and/or UV wavelengths in nearly every stellar 
rotation and occasionally they produce very large flares.  

Just like BO~Mic and AB~Dor, the star LO~Peg (BD +22 4409) is also
a very young low mass rapid rotator with a spectral type between K3 and K8 
\citep{jefferies_1994, pandey_2005}. 
LO~Peg is located at a distance of $\sim$ 24~pc  and thought to be 
a member of the local association \citep{jeffries_1993, montes_2001}. 
On the basis of its galactic space motion and its large fractional EUV-Luminosity 
of $\frac{L_{EUV}}{L_{bol}}\sim$-3.53, \cite{jefferies_1993} 
identify LO~Peg as a member of the Pleiades moving group with an age between 
20~-~150~Myr, 
while \cite{zuckerman_2004} identify LO~Peg as a member of \textbf{AB~Dor} group of $\sim$ 50~Myr 
old stars in the solar 
neighbourhood.  \cite{jefferies_1994} suggest a rotation period of LO~Peg between 0.3841-0.42375 day
from V-band photometry. The same authors also find 
an equatorial rotational velocity $v~sini$ $\sim$~69$\pm$1~km/s.  

LO~Peg was detected as a X-ray source in the ROSAT
all-sky survey (RASS) as the source 1RXS J213101.3+232009 \citep{voges_1999} with
an X-ray luminosity (in the 0.1~-~2.0~keV energy band) of 5.1 $\times$ 10$^{29}$ $erg/s$
and a ratio of X-ray to bolometric luminosity of $\frac{L_X}{L_{bol}}\approx-3.2$.
LO~Peg was also detected with the \emph{ROSAT} Wide Field camera (WFC) as the source
RE~2131~+23.3 and with the
Extreme Ultra-Violet Explorer (EUVE) as the source  EUVE~2131+233  \citep{malina_1994};
the coarse X-ray spectral properties of LO~Peg were explored with a short
ROSAT PSPC observation by \cite{pandey_2005}.   All the available data
suggest that LO~Peg does indeed have a powerful corona 
close to the saturation limit indicating that the star is magnetically very active.

The magnetic activity of LO~Peg was also demonstrated by its strong H$\alpha$ 
and \ion{Ca}{ii} H\&K emission lines  \citep{jefferies_1994}. 
Furthermore, \cite{eibe_1999} presented evidence for optical 
flaring on LO~Peg and an intense down-flow of material.  
The rapidly changing surface activity and high equatorial rotational velocity 
make LO~Peg an interesting object in terms of stellar activity and its relation
to stellar rotation. 
Since the discovery of LO~Peg as a variable star, it has been the  subject of detailed studies like spectral 
surface mapping and long-term photometric observations. 
For example, \cite{tas_2011} showed that LO~Peg has an active longitude of 
lifetime $\sim$~1.3 years  and $\sim$~4.8~yrs activity cycle period. 

In order to provide a more detailed picture of the coronal properties of
this ultra-fast rotator, we have obtained XMM-Newton data 
covering a full rotation period of LO~Peg, 
which we present and discuss in this paper.  Our paper is structured as follows:
In Section~2 we describe our observations obtained with  {\em XMM-Newton}. 
In Section~3 and 4, we characterise  the temporal and spectral behaviour of LO Peg. 
The coronal properties of LO~Peg are compared with other rapid rotators in Section~5 and in 
Section~6 we present our summary and conclusions.

\section{Observations and data analysis}

The new X-ray data on LO~Peg presented in this paper have been obtained with 
the XMM-Newton satellite on November, 30$^{th}$ 2014 (PI Lalitha S.; Obs ID: 0740590101).
Onboard the XMM-Newton satellite there are three co-aligned X-ray telescopes \citep{jansen_2001} and the 
European Photon Imaging Camera (EPIC; see \citealt{struder_2001, turner_2001}), which contains
three CCD cameras (one pn and two MOS cameras) with a sensitivity range in the
energy band $\sim$ 0.2~-~15~keV.  The EPIC X-ray CCD detectors provide medium-resolution imaging 
spectroscopy E/$\delta$E $\sim$~20~-~50 and temporal resolution at the sub-second level. 
The X-ray telescopes with the MOS detectors are also equipped with 
simultaneously operated reflection gratings (the so-called 
Reflection Grating Spectrometer -- RGS; \citealt{den_2001}), which provide 
high-resolution X-ray spectroscopy in the energy range 0.35~-~2.5 keV.  Finally,
XMM-Newton carries an Optical Monitor (OM), i.e., an optical/UV telescope with 
different filters for imaging and time-resolved photometry \citep{mason_2001}. 

\begin{figure}
\begin{center}
\includegraphics[width=9cm,height=12cm,clip]{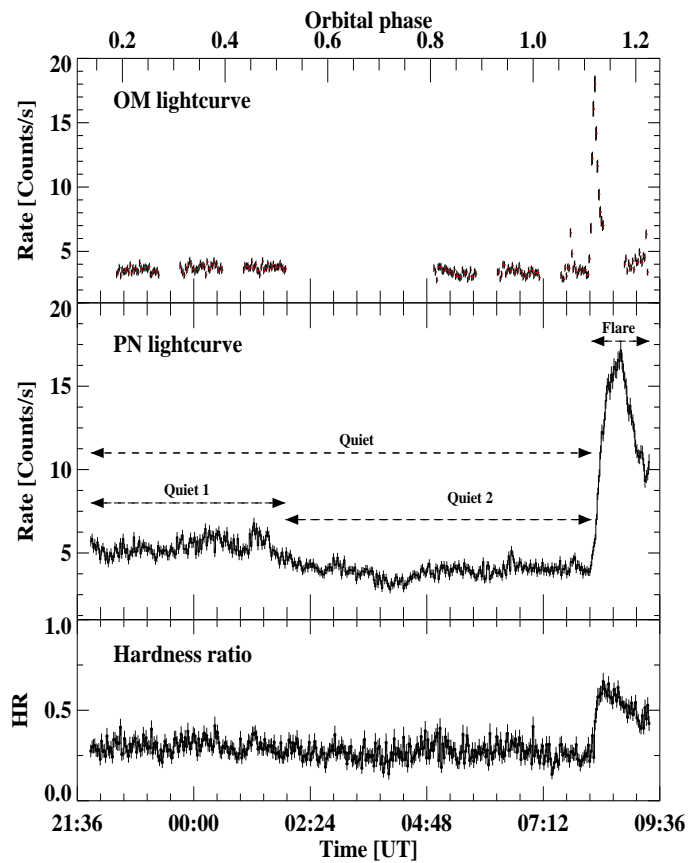}\vspace{-0.4mm}
\caption{\label{light_curve} \emph{XMM-Newton} light curves and hardness ratios of LO~Peg obtained 30 Nov 2014, taken simultaneously by the OM and EPIC detectors and binned to 100~s. The arrows represent the different time bins used for further analysis; see text for details.}
\end{center}
\end{figure}

Our XMM-Newton observations have a total duration of 42~ks, covering more than one rotation 
period of LO~Peg (P$_{rot} \sim$ 36 ks).   Useful data for LO~Peg are available from 
the OM, EPIC, and the RGS detectors.
The pn and MOS detectors were operated with the medium filter in imaging and small window mode, while the OM was operated in fast mode with a 0.5~sec cadence using the UVM2 band filter covering a band pass between 205~-~245~nm.

All X-ray data were reduced with the XMM-Newton Science Analysis System (SAS) 
\footnote{The \emph{XMM-Newton} SAS user guide can be found at http://www.cosmos.esa.int/web/xmm-newton/documentation/} software, version
13.0. EPIC light curves and spectra were obtained using standard filtering criteria, spectral analysis was done with XSPEC version 12.8.1. \citep{arnaud_1996} 
for the overall fitting processes.  For model fitting we always assume a collisionally ionised optically thin gas as calculated with the Astrophysical Plasma Emission Code (APEC)   \footnote{http://hea-www.harvard.edu/APEC/};  \citealt{smith_2001}) and abundances are calculated relative to the solar photospheric values 
of \cite{grevesse_1998}.

\begin{figure}
\begin{center}
\includegraphics[width=8.5cm,clip]{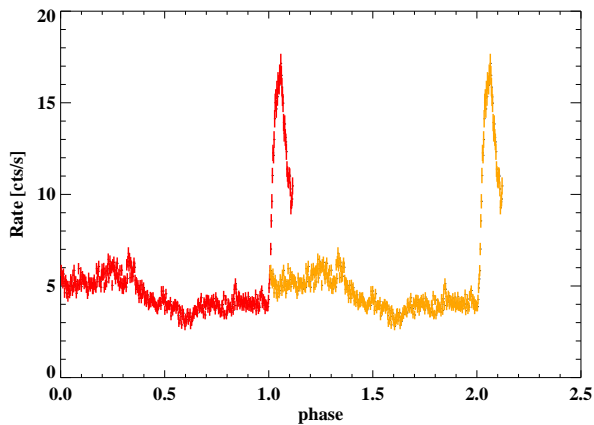}\vspace{-0.4mm}
\caption{\label{fig:period} The \emph{XMM-Newton} pn data, plotted in Fig.~\ref{light_curve},
folded with rotation period and plotted vs. the phase interval [0,2.0].  Note that
every data point is plotted twice; the original data is shown in red, the 
repeated data is shown in orange curve.}
\end{center}
\end{figure}

\section{Results}
\subsection{Timing analysis of XMM-Newton data: Overview}

For an overview of our XMM-Newton data we plot in Fig.~\ref{light_curve} the light curves 
simultaneously recorded with 
the OM (top panel) and the EPIC-pn (middle panel) detectors, each binned to 100~s;
the arrows in the plot show different time bins used for our later
spectral analysis.  As is evident from Fig.~\ref{light_curve},
a large flare occurred towards the end of observation when
the count rate increased from quiescent values of $\sim$ 4 cts $s^{-1}$ to 17 cts $s^{-1}$ in the pn 
detector and from $\sim$ 1 cts $s^{-1}$ to 5 cts $s^{-1}$ in the MOS detectors. 

To give a coarse description
of the changes in the spectral behaviour of the X-ray emission
we compute a hardness ratio (HR) for the pn as the ratio of the number of counts in hard band (2.0-10.0 keV) and the soft band (0.2-1.0 keV), which is plotted 
in the lower panel of Fig.~\ref{light_curve}.
We note that the HR values seem to be more or less constant at a level of $\approx 0.25 $ 
without significant variation during the quiescent phase, while 
a clear hardening to $\approx 0.75$ is seen during the large flare. 

\subsubsection{Rotational modulation~?}

In addition to the flare observed toward the end of our observations, LO~Peg 
also shows some modulations in the overall light curve. In Fig.~\ref{fig:period}
we plot the pn light curves folded with the ephemeris 
by \cite{dal_2003} HJD 2,448,869.93 + 0.42375 $\times$ E; we show our data twice, first in the
original form and then shifted by exactly one rotational period.
As shown in Fig.~\ref{fig:period}, our observations
cover hardly more than one rotation period and the large flare occurred at the very end so
that no real phase overlap between adjacent orbits is available in our data. 
Yet it appears that at the end of the orbit the count rate level is approximately the same
as at the beginning and it is tempting to interpret the apparent light curve modulation as 
a rotation-induced modulation. 
The light curve reaches its maximum at phase $\phi \approx $ 0.25-0.3, and
the large flare took place between phases 1~-~1.1, consistent with the interpretation
that a more active hemisphere could have been in view. 

Assuming that the quiescent emission of LO~Peg exhibits partial rotational modulation, we can estimate the 
degree of modulation using the amplitude of a simple sine wave fit to the data.  Using only
the quiescent data, we find an amplitude of 0.45~$\pm$~0.01~cts~$s^{-1}$ and a mean level 
of 3.39~$\pm$~0.17~cts~$s^{-1}$, corresponding to a rotational modulation of 14$\%$.

\begin{figure}
\begin{center}
\includegraphics[width=8cm,clip]{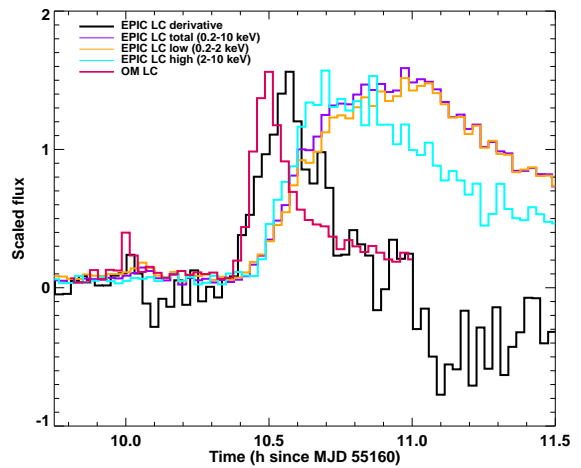}\vspace{-0.4mm}
\caption{\label{neupert} Close-up view of the large flare observed on LOPeg. Depicted are the combined EPIC X-ray light curve in 0.2-10 keV energy band in violet and its time derivative (smoothed by five bins) in black. We also plot the EPIC X-ray soft (0.2-2 keV in orange) and hard (2-10 keV in cyan) light curves along with the OM light curve (in pink).}
\end{center}
\end{figure}

\subsubsection{Neupert effect}

\begin{figure*}
\begin{center}
\includegraphics[width=0.9\textwidth]{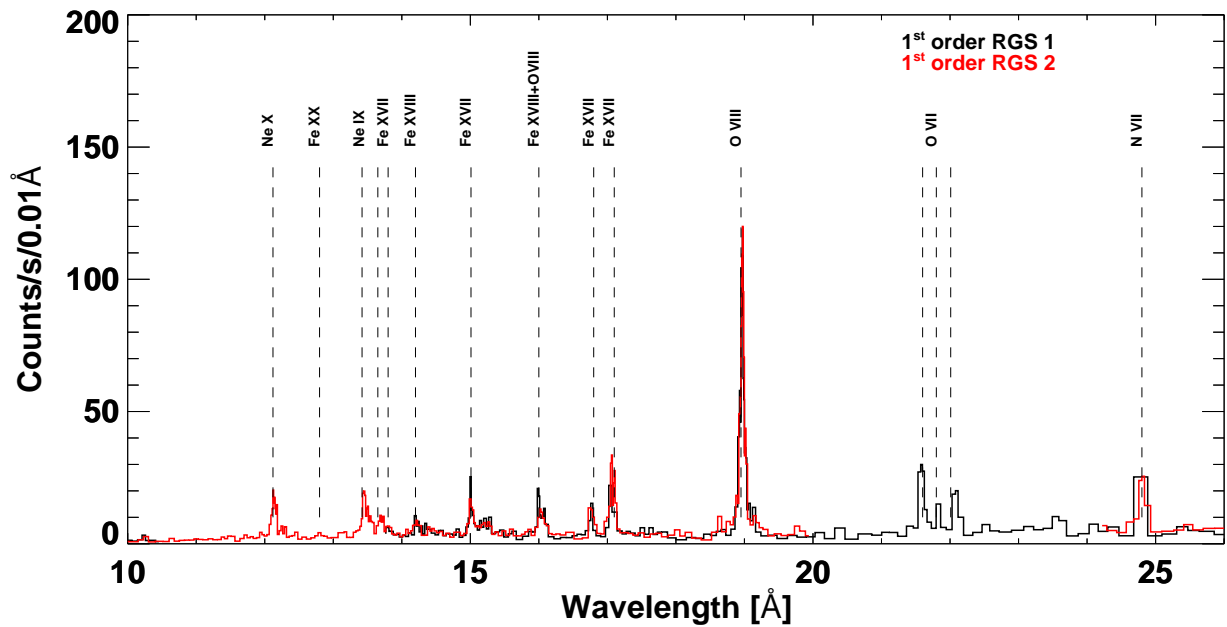}
\caption{\label{fig:rgs_spec} Quiescent time-integrated 1$^{st}$ order RGS 1 (black data points)  
and RGS 2 (red data points) spectra of LO Peg in the 10-26 $\AA$ wavelength ranges with 
the strongest lines labelled.}
\end{center}
\end{figure*}
In Fig.~\ref{neupert} we show a close-up view of the LO~Peg flare light curve in various
bands, i.e., in the ``soft'' X-ray band (between 0.2-2 keV; orange), in the ``hard''
X-ray band (between 2-10 keV; cyan) and in the total X-ray band (between 0.2-10 keV; violet).
We furthermore plot the time derivative of soft X-ray EPIC light curve (black) as well as the
OM light curve (red), all the lightcurves are binned to 100~s and shifted to be around zero and
scaled in amplitude to be as consistent as possible with the time derivative ``light'' curve.
A visual inspection of Fig.~\ref{neupert} clearly indicates the peak of the harder X-ray band (cyan curve)
precedes the peak of the softer X-ray band (orange curve) by $\sim$~800-1000s, while the OM light curve peak (red) precedes the
peak of the harder X-ray band (cyan curve) by about 15 minutes or 900~s; the shape of the time derivative of the
soft X-ray EPIC light curve (black) agrees well, albeit not perfectly, with the OM light curve, with a peak delay
of 2 bins or 400~s.  This phenomenology suggests to interpret the OM emission as a proxy for the expected non-thermal radiation 
as a result of electron impingement on the chromosphere of LO~Peg, which leads to the observed (thermal) X-ray flare, and thus 
as evidence for the Neupert effect.

We investigated further evidence for the Neupert effect during the rise phase of the flare. 
If, during a flare, the energy contained in all the accelerated particles is 
used for chromospheric and coronal heating, 
the time integral of the emission due to the accelerated particles or one of its proxies
like hard X-ray emission, white-light emission, etc. resembles the 
rise of the flare light curve in the soft X-ray band \citep{neupert_1968}. 
This can be represented as 
\[
 F_{SXR} \propto  \int F_{HXR}(t) dt     ~~~~ or ~~~~ \frac{d}{dt} F_{SXR}(t) \propto F_{HXR}(t), 
\]
where F$_{SXR}$ and F$_{HXR}$ represent the soft and hard (non-thermal) X-ray fluxes, respectively.

Since there is no HXR data for our observations we use the OM data as a proxy for the 
emission due to the accelerated particles. In Fig.~\ref{neupert}, we also plot the time derivative of soft X-ray EPIC light curve in black along with the ultra-violet band (optical monitor in pink). 
We note that the EPIC light curve during flare rise matches the shape of the OM light curve.

Cross-correlating the OM~light curve and the time derivative of the X-ray light curves, we obtain a time-lag between the two peaks to be  $\sim$~250s, using the Z-transformed Discrete Correlation 
Function (ZDCF) technique described by \cite{tal_1997}, much less than the delay between the
soft X-ray and OM peak.  This strongly suggests that indeed
the optical/UV peak is a good proxy of the emission due to accelerated particles,
preceding the bulk of the soft X-ray emission.  

\subsection{Spectral analysis of XMM-Newton data: Overview}

Next, we examine the EPIC and RGS spectra of LO Peg extracted during flaring and quiescent intervals
to study plasma temperatures, emission measures and abundances change as a result of flaring. 
In Fig.\ref{fig:rgs_spec}, we 
plot the quiescent time averaged 1$^{st}$ order RGS 1 (black data points) and RGS 2 (red points) in
the 10-26 $\AA$ wavelength range. The strongest lines are identified and labelled and are -- as expected --
due to oxygen, neon, iron and nitrogen.

\begin{figure}
\begin{center}
\includegraphics[width=8cm]{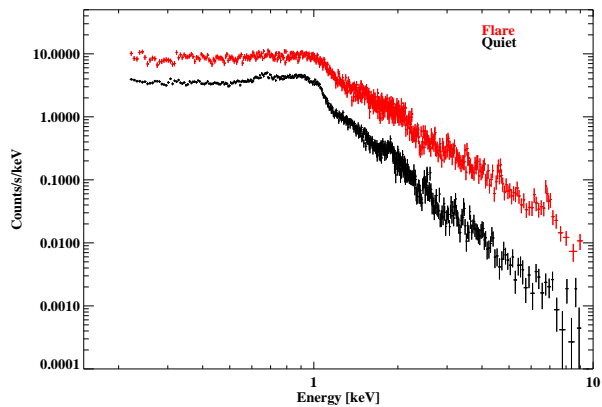}
\caption{\label{fig:spectra} EPIC PN spectra for different time-bins quiet (black) and the flare (red).}
\end{center}
\end{figure}

During the course of a flare, fresh material from the chromosphere is heated, evaporated and 
transported into the corona, thus temporarily changing the coronal emission measure and 
possibly the coronal abundances.  
In Fig.~\ref{fig:spectra}, we plot the EPIC pn spectra extracted during quiescence (in black) and the flare (in red). 
The flare-related changes in the spectral energy distribution are very much 
evident; the spectral hardening is already apparent in the lower panel of Fig.~\ref{light_curve} depicting the hardness ratios.

Along with the flare related changes, we also examine possible orbital 
variations by dividing the overall quiescent spectra in periods dubbed as 
``Quiet 1'' and ``Quiet 2'' (the time bins for the spectra are as shown in Fig.~\ref{light_curve}) 
and model the pn, MOS and RGS X-ray spectra. 
We specifically determine the temperatures, emission measures and abundances relative to solar values \citep{grevesse_1998} with simultaneous iterative 
global XSPEC fits to the combination of EPIC and RGS (RGS+PN or RGS+MOS) spectra with variable-APEC (VAPEC; \citealt{smith_2001}) plasma models. 
As is often observed \citep{gudel_2001, robrade_2005, lalitha_2013}, we require
multi-temperature components to achieve an adequate description of the observed coronal spectra.  
We use combinations of two, three and four temperature components and find that
a three temperature component model leads to an adequate description of the data. 
We fit each of these spectra in the full 0.2~-~10~keV energy range. 
For fitting the RGS spectra, the temperature and the abundances of elements like 
carbon, nitrogen, oxygen, neon and iron are allowed to vary freely and independently, 
however, the abundances are fixed among the different APEC temperature components. 
For fitting EPIC-MOS or pn spectra we allow the magnesium, sulphur, and silicon abundances 
to vary along with the oxygen, neon and iron abundance. However, the carbon and nitrogen 
abundances are fixed to values obtained from the RGS, which is more sensitive to strong 
individual lines of these elements. 
In Tab.~\ref{tab:fit_param}, we summarise the results of this fitting procedure along 
with the 90\% confidence range errors.  Tab.~\ref{tab:fit_param} shows that the
quiescent state is characterised by dominant plasma components at $\sim~3$, $\sim~7.5$ and $\sim~20$~MK, while during the flare the coronal temperature bins increase to  $\sim~3.5$, $\sim~12$ and $\sim~32$~MK. 
During the flare, a pronounced enhancement of the emission measure at $2.8$~keV is present,
indicating the rise of emission measure at a higher temperature. The quiescent bins (quiet 1 and quiet 2) does not show any significant difference in the coronal properties when compared to the overall quiescent time-bin; suggesting no changes in coronal properties with orbital variation.

\begin{table*}
\caption{\label{tab:fit_param} Results on temperature, emission measures and abundance relative to solar photospheric values.}
\begin{tabular}[htbp]{ccccccccc}
\hline
\hline
Param & \multicolumn{2}{c}{Quiet} & \multicolumn{2}{c}{Quiet1} & \multicolumn{2}{c}{Quiet2} & \multicolumn{2}{c}{Flare} 	\\

 &     RGS+PN      & RGS+MOS &     RGS+PN      & RGS+MOS&     RGS+PN      & RGS+MOS&     RGS+PN      & RGS+MOS\\
\hline
kt1 [keV] &       0.26$_{-0.01}^{+0.01}$ & 0.27$_{-0.01}^{+0.02}$ &0.26$_{-0.01}^{+0.01}$   &  0.27$_{-0.01}^{+0.02}$   &     0.26$_{-0.01}^{+0.01}$   & 0.26$_{-0.01}^{+0.02}$ &  0.30$_{-0.02}^{+0.02}$   &     0.32$_{-0.02}^{+0.07}$ \\  
EM1 [10$^{52}$ cm$^{-3}$] &       1.16$_{-0.09}^{+0.10}$ & 0.94$_{-0.07}^{+0.08}$ &1.14$_{-0.17}^{+0.18}$   &  1.06$_{-0.12}^{+0.13}$   &     1.15$_{-0.12}^{+0.13}$   & 0.93$_{-0.10}^{+0.11}$ &  2.24$_{-0.45}^{+0.50}$   &     1.21$_{-0.31}^{+0.36}$ \\  
kt2 [keV]&       0.65$_{-0.01}^{+0.01}$ & 0.66$_{-0.01}^{+0.01}$ &0.64$_{-0.02}^{+0.02}$   &  0.66$_{-0.02}^{+0.04}$   &     0.66$_{-0.02}^{+0.02}$   & 0.66$_{-0.02}^{+0.01}$ &  1.00$_{-0.04}^{+0.03}$   &     1.00$_{-0.07}^{+0.05}$  \\ 
EM2 [10$^{52}$ cm$^{-3}$]&       1.99$_{-0.17}^{+0.29}$ & 1.70$_{-0.14}^{+0.15}$ &2.62$_{-0.60}^{+0.37}$   &  1.86$_{-0.23}^{+0.25}$   &     1.73$_{-0.20}^{+0.22}$   & 1.64$_{-0.18}^{+0.19}$ &  2.90$_{-0.63}^{+0.75}$   &     1.82$_{-0.45}^{+0.52}$ \\  
kt3 [keV]&       1.66$_{-0.06}^{+0.17}$ & 1.56$_{-0.04}^{+0.05}$ &1.81$_{-0.24}^{+0.14}$   &  1.58$_{-0.05}^{+0.06}$   &     1.62$_{-0.10}^{+0.11}$   & 1.52$_{-0.06}^{+0.07}$ &  2.87$_{-0.21}^{+0.22}$   &     2.81$_{-0.20}^{+0.18}$ \\  
EM3 [10$^{52}$ cm$^{-3}$]&       1.29$_{-0.15}^{+0.10}$ & 1.43$_{-0.07}^{+0.07}$ &1.58$_{-1.62}^{+0.31}$   &  2.00$_{-0.13}^{+0.13}$   &     1.00$_{-0.10}^{+0.96}$   & 1.08$_{-0.09}^{+0.08}$ &  7.73$_{-0.55}^{+0.60}$   &     7.85$_{-0.48}^{+0.57}$ \\  
C   &       0.60$_{-0.17}^{+0.18}$ & 0.51$_{-0.20}^{+0.21}$ &0.78$_{-0.46}^{+0.52}$   &  0.16$_{-0.15}^{+1.11}$   &     0.45$_{-0.28}^{+0.30}$   & 0.31$_{-0.30}^{+0.35}$ &  0.89$_{-0.85}^{+0.80}$   &     1.13$_{-1.13}^{+1.82}$\\
N   &       0.53$_{-0.14}^{+0.15}$ & 0.65$_{-0.17}^{+0.19}$ &0.66$_{-0.27}^{+0.30}$   &  0.76$_{-0.30}^{+0.33}$   &     0.48$_{-0.19}^{+0.20}$   & 0.55$_{-0.22}^{+0.24}$ &  0.54$_{-0.54}^{+0.51}$   &     1.64$_{-1.08}^{+1.40}$\\
O   &       0.38$_{-0.02}^{+0.03}$ & 0.46$_{-0.03}^{+0.04}$ &0.39$_{-0.05}^{+0.05}$   &  0.47$_{-0.04}^{+0.05}$   &     0.36$_{-0.03}^{+0.04}$   & 0.41$_{-0.04}^{+0.04}$ &  0.41$_{-0.08}^{+0.11}$   &     0.74$_{-0.17}^{+0.21}$\\
Ne  &       1.17$_{-0.07}^{+0.08}$ & 1.37$_{-0.09}^{+0.10}$ &1.19$_{-0.13}^{+0.15}$   &  1.38$_{-0.14}^{+0.15}$   &     1.14$_{-0.10}^{+0.11}$   & 1.28$_{-0.12}^{+0.13}$ &  0.81$_{-0.23}^{+0.27}$   &     1.24$_{-0.36}^{+0.44}$\\
Mg  &       0.26$_{-0.04}^{+0.05}$ & 0.31$_{-0.05}^{+0.05}$ &0.21$_{-0.07}^{+0.08}$   &  0.30$_{-0.07}^{+0.08}$   &     0.29$_{-0.06}^{+0.07}$   & 0.33$_{-0.06}^{+0.07}$ &  0.11$_{-0.11}^{+0.10}$   &     0.31$_{-0.22}^{+0.25}$\\
Si  &       0.36$_{-0.05}^{+0.05}$ & 0.28$_{-0.04}^{+0.05}$ &0.37$_{-0.07}^{+0.08}$   &  0.27$_{-0.06}^{+0.07}$   &     0.33$_{-0.06}^{+0.07}$   & 0.26$_{-0.06}^{+0.06}$ &  0.44$_{-0.14}^{+0.15}$   &     0.46$_{-0.16}^{+0.17}$\\
S   &       0.23$_{-0.06}^{+0.07}$ & 0.31$_{-0.06}^{+0.07}$ &0.24$_{-0.10}^{+0.11}$   &  0.32$_{-0.10}^{+0.10}$   &     0.25$_{-0.09}^{+0.10}$   & 0.33$_{-0.09}^{+0.10}$ &  0.12$_{-0.12}^{+0.10}$   &     0.25$_{-0.19}^{+0.19}$\\
Fe  &       0.20$_{-0.01}^{+0.02}$ & 0.23$_{-0.02}^{+0.02}$ &0.18$_{-0.02}^{+0.03}$   &  0.23$_{-0.02}^{+0.03}$   &     0.22$_{-0.02}^{+0.03}$   & 0.23$_{-0.02}^{+0.02}$ &  0.34$_{-0.06}^{+0.07}$   &     0.50$_{-0.09}^{+0.10}$\\
\hline
$\chi^2$ &  1.23     &   1.22       & 1.14     &   1.04     &  1.23     &  1.22    &  1.01    &   0.97\\
DOF      & 1273        &   1136       & 704        &   585        &   735 	& 615	    &  704        &    571\\
\hline
L$_X$ [erg/s]    & 29.69   &   29.68 &     29.77 &      29.76 &     29.62  &     29.62 &      30.21   &   30.20 \\
\hline
\hline
\end{tabular}

\footnotesize{Note: the errors are estimated with 90\% confidence limit. L$_X$ indicates the 0.2-10 keV luminosity. }\\

\end{table*}

 \begin{figure}
\begin{center}
\includegraphics[width=8cm]{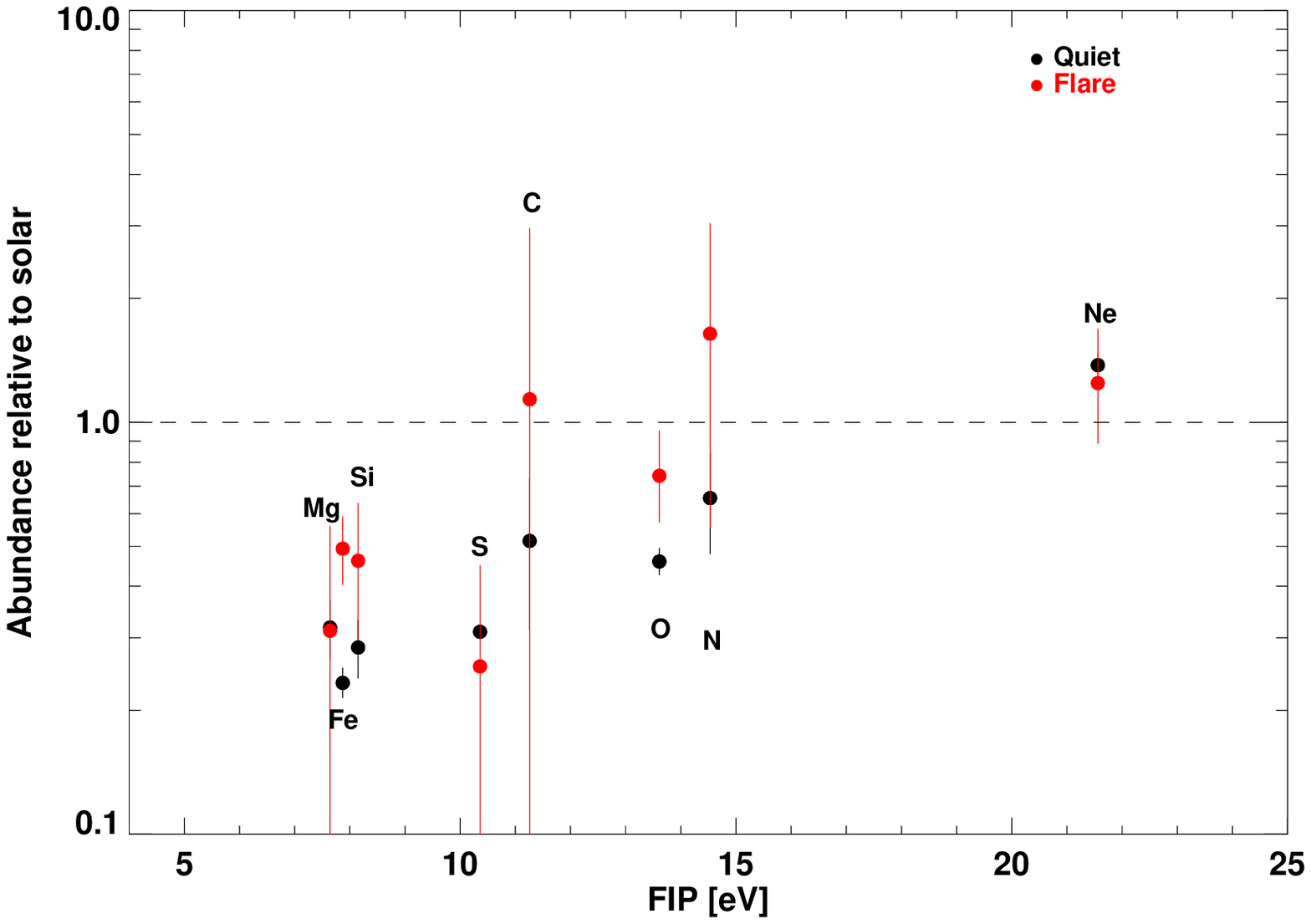}
\includegraphics[width=8cm]{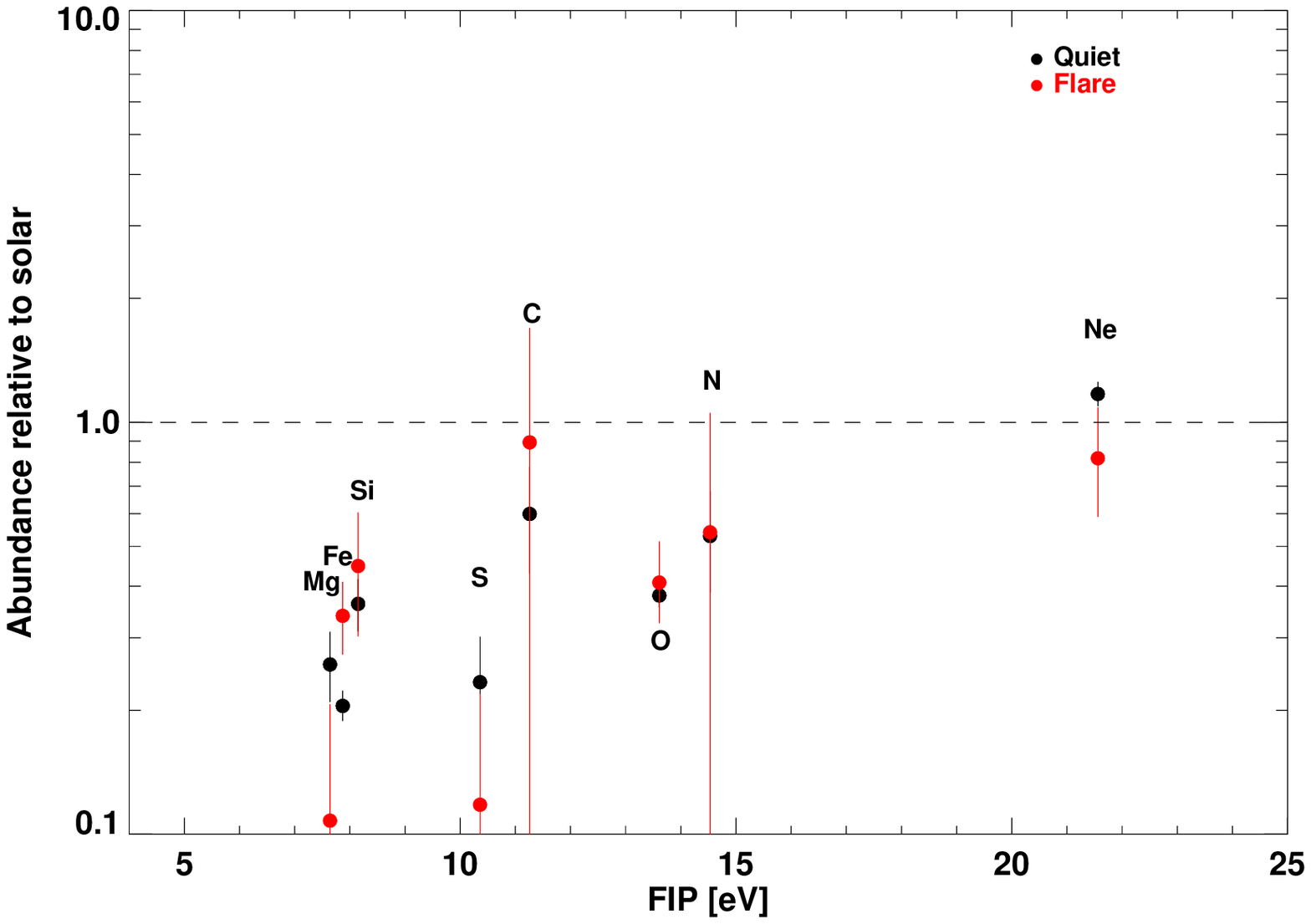}
\caption{\label{fig:fip} Elemental abundances relative to solar photospheric values \citep{grevesse_1998} as a function of the first ionisation potential (FIP) during quiescence (black) and flare (red). The top and bottom panel shows abundance values obtained using  a combination of MOS+RGS and PN+RGS, respectively.  
Dashed line at unity indicates the solar photospheric abundance.}
\end{center}
\end{figure}

\begin{figure}
\begin{center}
\includegraphics[width=0.5\textwidth]{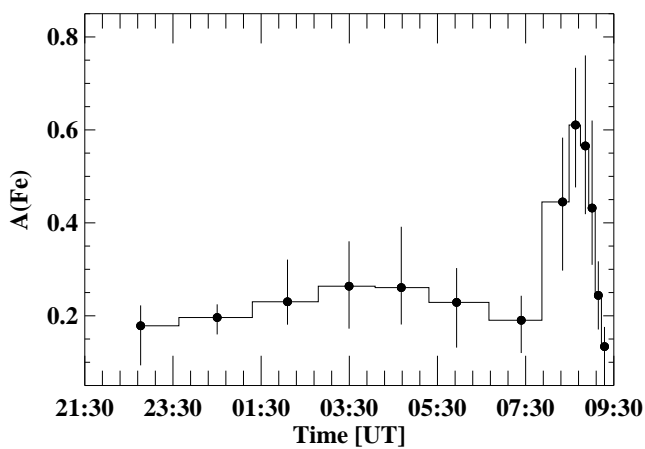}
\caption{\label{fig:feevolve} Temporal evolution of the Fe abundance as obtained from the EPIC-pn data.}
\end{center}
\end{figure}

The abundance patterns of LO~Peg during different activity states are shown in 
Fig.~\ref{fig:fip}, where we plot the abundances with respect to solar photospheric abundances against the FIP (First Ionisation Potential) of the corresponding element. The top and bottom panels show the abundance patterns obtained for RGS+PN and RGS+MOS combinations, respectively,  for the quiescent and the flaring time bins. Inactive stars like the Sun show the FIP effect, where elements with low-FIP elements like Fe, Si, and Mg are enhanced in the corona when compared to high-FIP elements like C, N, O, and Ne. A reversed pattern, 
the inverse FIP effect with 
enhanced high-FIP elements when compared to the low-FIP elements, is frequently observed in active stars (e.g., \citealt{brinkman_2001}, \citealt{audard_2003}). The abundance pattern of 
LO~Peg indicates an inverse FIP effect as can be seen in Fig.~\ref{fig:fip}. 
Furthermore, during the flare some of the abundances such as Fe, O, and Si,
are found to be slightly higher (albeit with uncomfortably large errors), and the Fe and Si 
abundances also rise compared to the high-FIP elements.
In Figure\ref{fig:feevolve}, we plot the evolution of Fe abundance showing a significant increase from the quiescence level before the flare to a maximum during flare peak and then decreases to the pre-flare values. All this is consistent with the picture that the fresh chromospheric material evaporates into the corona, 
thus changing the elemental abundances. 
We note that the global fit to the data using APEC models actually measures the absolute abundance constrained by the flux in the line (which
is proportional to the abundance of the element producing the line) and
the overall fit to the continuum, which is dominated by the hydrogen and helium
content of the plasma as the main donors of electrons.

\subsubsection{Emission measure distribution}
 
In contrast to the abundances, emission measures and temperatures of the coronal plasma change more significantly during the flare evolution. 
The plasma emission measure increases and the spectra hardens as the temperatures rise (see Fig.~\ref{fig:spectra} ). 
For instance, the Fe line complex at 6.7~keV, which is formed at 40~MK, appears only
during the flaring state.  

\begin{figure}
\begin{center}
\includegraphics[width=9cm]{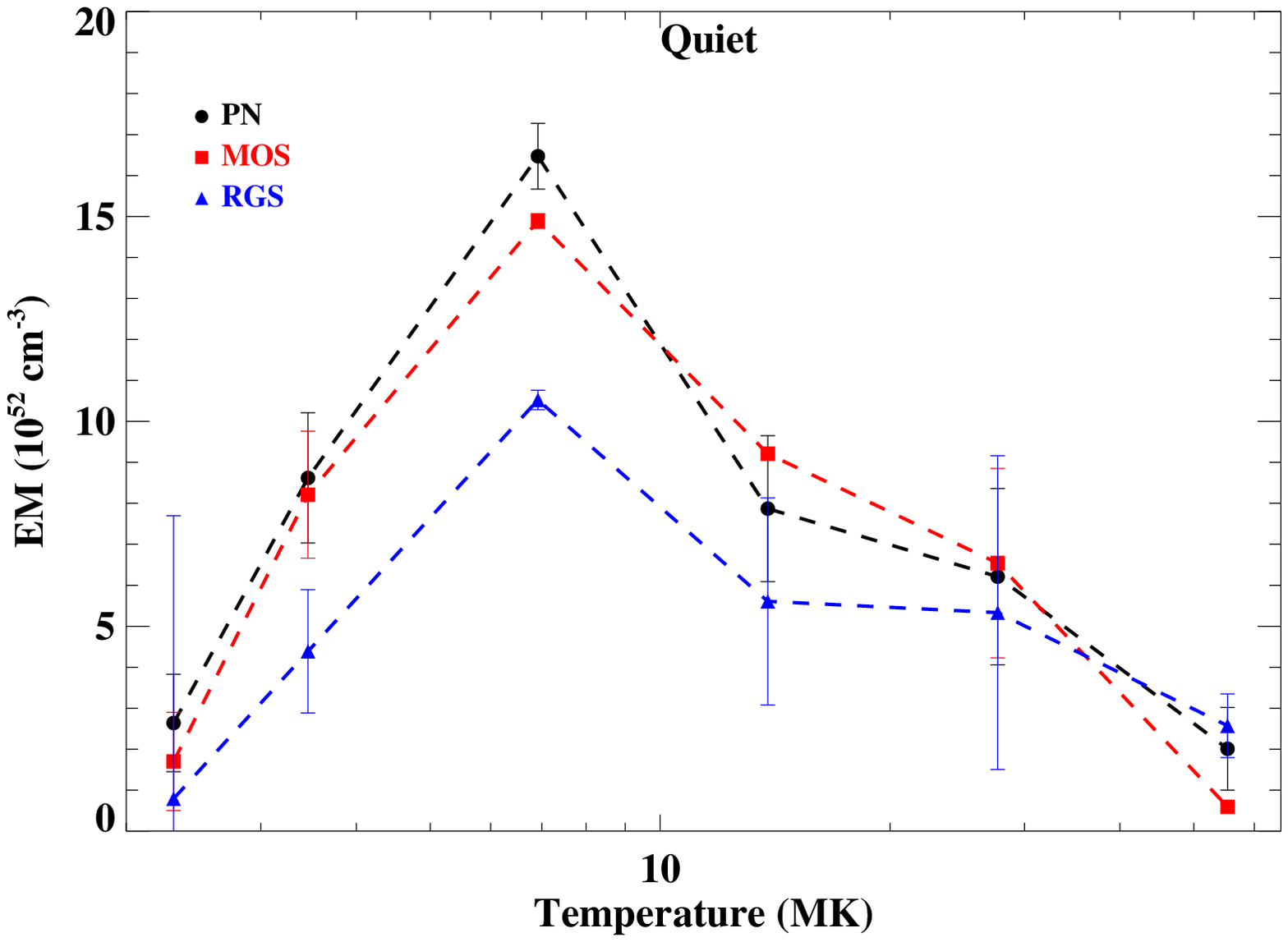}
\includegraphics[width=9cm]{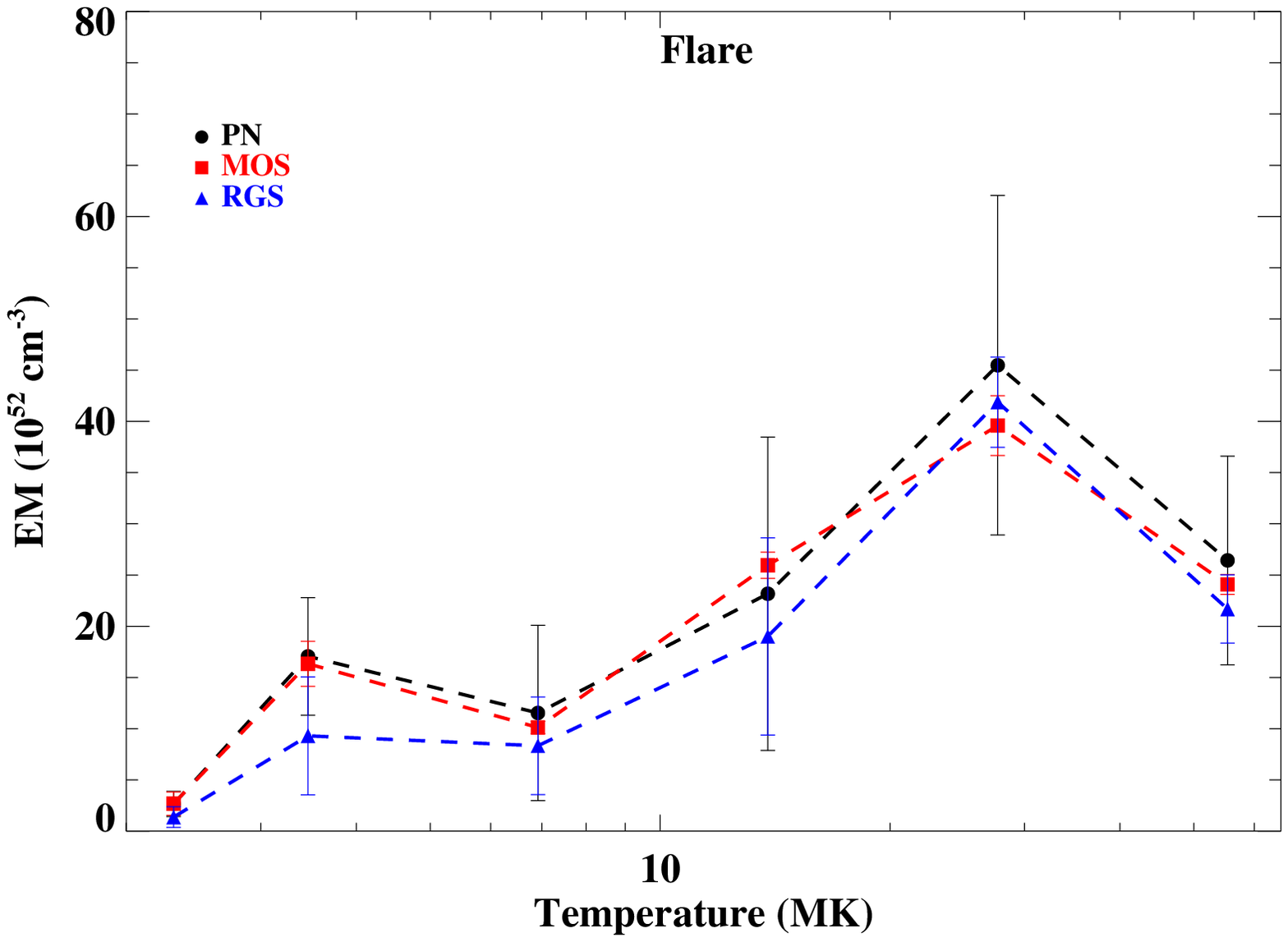}
\caption{\label{fig:emd} EMDs of LO~Peg during different states of activity derived from PN (black filled circles), MOS (red filled squares) and RGS (blue filled triangles) data. The upper panel shows quiescent phase and lower panel shows flaring state of LO~Peg.  }
\end{center}
\end{figure}

We carry out a quantitative study of the temporal evolution of the plasma temperature during different activity states by constructing the emission measure 
distribution (EMD). We use the pn, MOS and RGS data to study the changes in the 
EMD during different activity states. To construct an EMD we use 
a 6-T VAPEC model with a fixed temperature grid 
(put at energies of 0.2, 0.3, 0.6, 1.2, 2.4, 4.8~keV),  where the first two components 
represent cool plasma (2-5~MK), the third and fourth temperature components represent medium temperature plasma (5-20~MK) and the last two components represent the hot plasma (20-60~MK). The abundances of neon, magnesium, oxygen, silicon, sulphur and iron are allowed to vary freely and independently, but are constrained to be same among all the other VAPEC temperature components. 
A graphical representation of the results on the EMDs using pn (in black), MOS (in red)  and RGS (in blue) data for LO~Peg during different activity states is shown in Fig.~\ref{fig:emd}. 
During the quiescent state the EM peaks around 7~MK and the hot component is only marginally detected. However, during the flare we note a  peak at 3.5~MK and  a large amount of additional plasma at around 30 MK with the
EM rising by a factor of $\sim$~7 at 30~MK. There is little or no change in the cool plasma and the main influence due to the flare is the increased amount of high temperature emission measure.

\subsubsection{Coronal densities}
\label{densitiesne}

Using the XMM-Newton RGS spectra we investigate the electron densities of coronal plasma from the density-sensitive line ratios of 
forbidden to inter-combination lines of the helium-like triplets of \ion{O}{vii}, the
theory of which has been described in detail by \cite{gabriel_1969}. If the electron 
collision rate is sufficiently high, the ions in the upper level of the forbidden transition level do not return to the ground level radiatively, instead, the ions are collisionally moved to the upper level of the inter-combination transitions, from where they eventually decay radiatively to the ground state. Therefore, the resulting ratio of the forbidden to inter-combination line (f/i) are sensitive to density. 

\begin{table*}
\begin{center}
\caption{\label{oviitable} Measured X-ray counts in f, i and r lines, $f/i$ ratios and 
deduced coronal densities from the \ion{O}{vii}  and \ion{Ne}{ix} triplet in the RGS spectra during different time bins including flare.}
\begin{tabular}[htbp]{llcccc}
\hline
\hline
ion& line & Quiet & Quiet1 & Quiet2 & Flare\\
\hline \\[-3mm]
&resonance (r, 21.60 \AA) 	& 23.60$\pm$7.71 & 7.26$\pm$3.83 		& 15.23$\pm$4.99		& 8.27$\pm$4.00 \\
&inter-combination (i, 21.80 \AA) 	& 24.52$\pm$7.86 & 5.87$\pm$2.57 		& 17.50$\pm$7.38		& 3.69$\pm$2.10 \\
\ion{O}{vii}&forbidden (f, 22.10 \AA) 		& 25.24$\pm$5.99 & 9.07$\pm$4.13 		& 18.53$\pm$10.20 		& 5.89$\pm$2.57 \\

&r$\sim \frac{f}{i}$  &1.07$\pm$0.42 & 1.54$\pm$0.97 & 1.05$\pm$0.73 & 1.59$\pm$1.14 \\

&log(n$_e$) [cm$^{-3}$]&10.81$\pm$0.76&10.72$\pm$0.55&10.81$\pm$0.57&10.71$\pm$0.50\\

\hline
&resonance (r, 13.44 \AA) 	& 76.98$\pm$18.45 & 34.26$\pm$8.59 		& 46.67$\pm$9.45		& 19.42$\pm$5.49 \\
&inter-combination (i, 13.55 \AA) 	& 45.97$\pm$22.32 & 12.37$\pm$6.42 		& 25.82$\pm$7.88		& 3.40$\pm$1.83 \\
\ion{Ne}{ix}&forbidden (f, 13.69 \AA) 		& 56.37$\pm$23.23 & 26.11$\pm$8.05 		& 30.85$\pm$9.78 		& 8.22$\pm$3.99 \\

&r$\sim \frac{f}{i}$  &1.22$\pm$0.78 & 2.11$\pm$1.27 & 1.19$\pm$0.52 & 2.41$\pm$1.75 \\

&log(n$_e$) [cm$^{-3}$]&11.96$\pm$0.11&11.82$\pm$0.12&11.97$\pm$0.08&11.78$\pm$0.15\\

\hline
\end{tabular}
\end{center}
\end{table*}

\begin{figure*}
\begin{center}
\includegraphics[width=9cm]{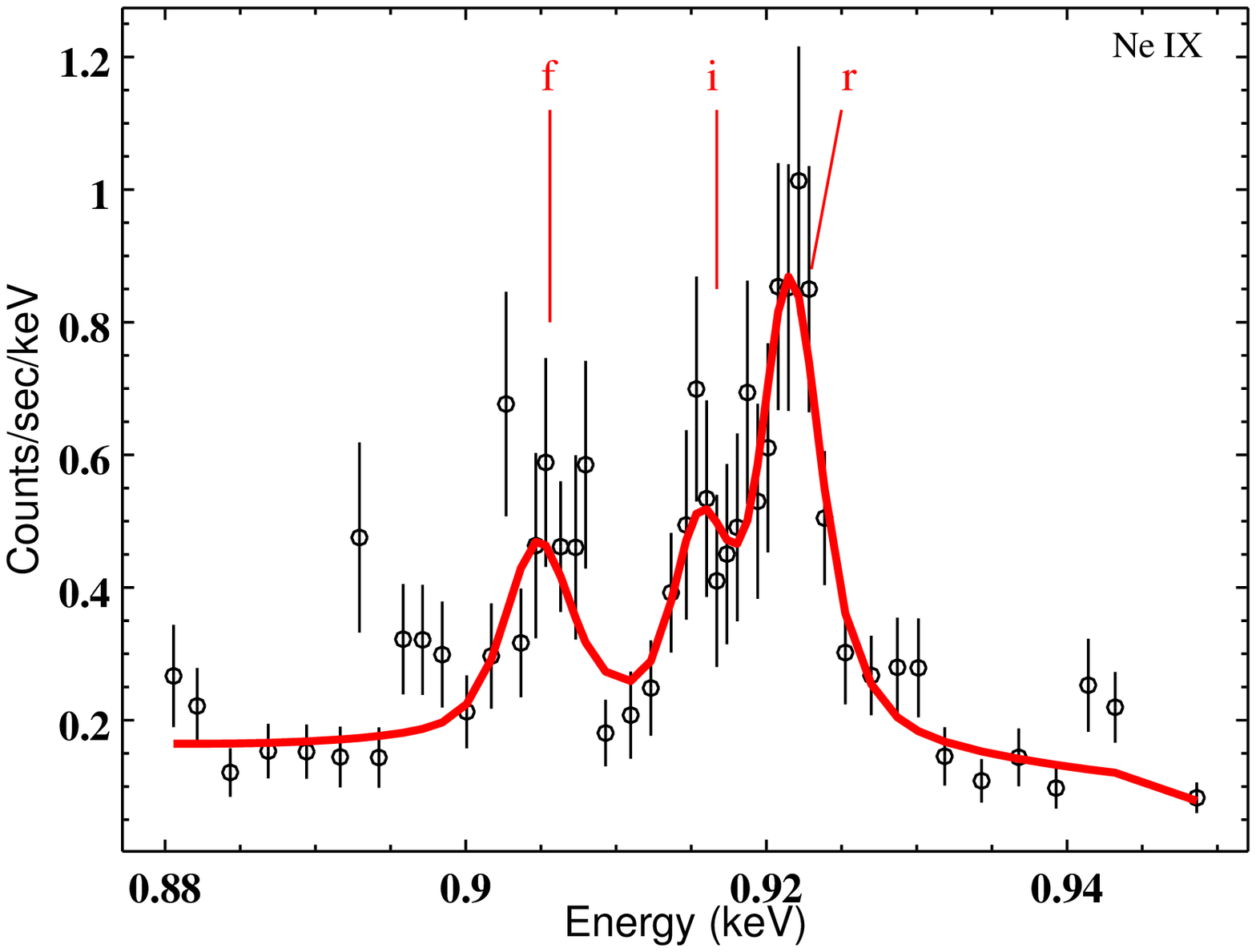}
\includegraphics[width=9cm]{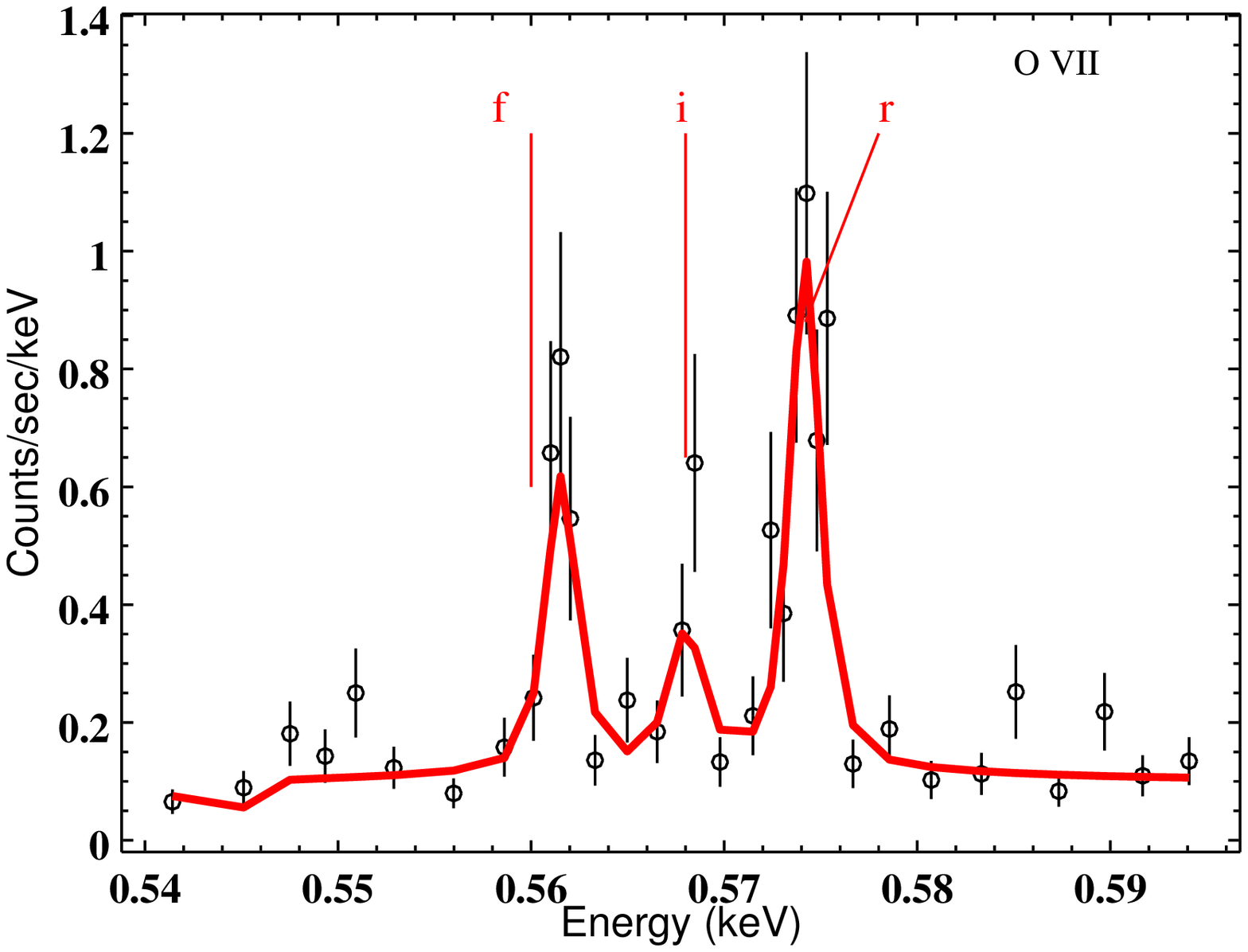}
\caption{\label{fig:den} Density-sensitive line triplet of \ion{Ne}{ix} (left panel) and \ion{O}{vii} (right panel) forbidden, inter-combination and resonance lies in increasing energies. Red line indicates the best-fit to the cumulated RGS data during quiescence.  }
\end{center}
\end{figure*}

The He-like triplet of \ion{Ne}{ix} and \ion{O}{vii} are strong enough in our observations
to be used to obtain characteristic electron densities in the
source region. 
The f, i and r lines for He-like triplets of Ne and O are fitted using Gaussian profiles to each of the line components.
In Fig.~\ref{fig:den}, we show the He-like triplets of Ne and O during the
quiescent state. 
The measured line counts, calculated
$f/i$ ratios and electron densities are listed in Tab.~\ref{oviitable}); 
note
that the quoted \ion{O}{vii} line counts are mean counts obtained from 
both the RGS1 and RGS2, while the \ion{Ne}{ix} triplet is recorded only by the RGS2.
To convert the measured f /i ratios to densities,
we use the expression
 \[
\frac{f}{i} = \frac {R_{o}}{1+\frac{n_{e}}{N_{c}}},
\]

\noindent where $R_{o}$ is 
the low-density limit and $N_{c}$ is the so-called critical density. We adopt the values from \cite{pradhan_1981} and use  $R_{o}$ = 3.95 and 
$N_{c}$ = 3.1 $\times$ 10$^{10}$~cm$^{-3}$ for
\ion{O}{vii}  and $R_{o}$ = 3.5 and $N_{c}$ = 5.9 $\times$ 10$^{11}$~cm$^{-3}$ for \ion{Ne}{ix}. 
The formal electron density calculated for LO~Peg using \ion{O}{vii} 
during quiescence and flare are log n$_e$[cm$^{-3}$]$\sim$10.81$\pm$0.76 and 10.71$\pm$0.50, respectively, while we find log n$_e$[cm$^{-3}$]$\sim$11.96$\pm$0.11 and 11.78$\pm$0.15, during quiescent and flaring states for \ion{Ne}{ix}.
Unfortunately, the number of measured counts in the f and i lines are 
rather small, leading to large errors.  While we can state that the \ion{Ne}{ix} emission is likely from a high density plasma region, we detect no significant differences between
the quiescent and flaring plasma, yet the errors are so large that we cannot
exclude an (expected) density increase in the flaring plasma. 
Finally, we note that the oxygen and neon triplet probe only temperatures 
between 1-4 MK and therefore the lines do not represent the major high-temperature
component of the flare.

\section{Discussion}

\subsection{Comparison with similar stars}

We now compare the derived coronal properties of LO~Peg 
with the properties of other ultra-fast rotators, i.e.,  AB~Dor, BO~Mic, PZ~Tel, HK~Aqr and YY~Gem. 
\begin{itemize}
\item[] a) AB~Dor is a well-studied example of an extremely active ultra-fast rotator (P$_{rot}$~=~0.51~d, d~$\sim$~15~pc, \citealt{guirado_2011}), which shows signatures  of activity at all wavelengths. 
In particular, AB~Dor shows strong photometric variability due to spots on its surface and 
is very bright in X-rays (L$_X$ $\sim$1$\times$10$^{30}$ erg/s, \citealt{lalitha_2013}).  

\item[] b)  BO~Mic is an highly active  young K-dwarf similar to AB~Dor. BO~Mic is also an ultra-fast rotator 
with a rotation period P$_{rot}$~=~0.38~d).  Its typical quiescent X-ray luminosity of is 
9 $\times$ 10$^{29}$ erg/s. Several flares are observed on BO~Mic, where
the flare energy can increase by up to two orders of magnitude \citep{kuerster_1995}. 

\item[] c) PZ~Tel is classified as a single K0V star 
with a rotation period of 0.94~d located at a distance 49~pc \citep{coates_1980, innis_1984, perryman_1997, zuckerman_2000}. PZ~Tel is a young post T-Tauri star (20 Myr, \citealt{favata_1998})  with an X-ray 
luminosity of  2.8$\times$10$^{30}$ erg/s \citep{stelzer_2000}. 
 
\item[] d) HK~Aqr is one of the fastest rotating M-dwarfs with a rotation period of 0.43~d ($v$ sin$i$ = 69 km s$^{-1}$, \citealt{young_1990}), located at a distance of $\sim$~22~pc. HK~Aqr is a single star with
a quiescent X-ray luminosity of 2$\times$ 10$^{29}$ erg/s.

\item[] e)  YY~Gem (Castor~C) is the faintest member of a triple system (Castor A+B+C) with an X-ray luminosity of  
2~-~8 $\times$ 10$^{29}$ erg/s, located at a distance of 15 pc.  YY~Gem itself is a spectroscopic
binary, in fact it is the brightest known eclipsing binary of the type dMe.
Both components are in a circular and synchronous orbit with an inclination angle of $\sim86^{\circ}$ \citep{pettersen_1976} and a period of 0.81~d \citep{kron_1952}, both components have 
have almost the same  spectral type.  The YY~Gem system is again magnetically very active star with a
high flare rate, dark cool star-spots from spectroscopic and photometric observations and X-ray variability (see \citealt{gudel_2001, hussain_2012}, and references therein).
\end{itemize}
We list the physical properties of our comparison stars along with those of LO~Peg in Tab.~\ref{prop}. 

\begin{table*}[!h]
\begin{center}
\caption{\label{prop} Comparison of the physical properties of LO~Peg with other ultra-fast rotating low mass stars.}
\begin{tabular}[htbp]{lcccccccccccc}
\hline
\hline
Star & Type & Dist& V&Age &v$ sin i$ & P$_{rot}$ & log L$_X$  & $\frac{L_X}{L_{bol}}$ \\
        &          & pc &mag& Myr& km/s   & days         &  erg/s         &                                              \\ 
\hline
AB Dor & K1 V  & $\sim$15 & 6.99 & 50& 90 & 0.51 & 29.93 & -3.25  \\
BO Mic & K0 V &  $\sim$44& 9.33 & 30 &132 & 0.38 & 29.94 &-3.07         \\
HK Aqr & M0 Ve&  $\sim$22& 10.87 &200 &70 & 0.44 & 29.24& -3.18     \\
YY Gem & M1  &  $\sim$15 & 9.83 &350& 37 ? & 0.81 & 29.39& -2.95  \\ 
PZ Tel &G9 IV  &  $\sim$50 & 8.34 &20&70 & 0.94   & 30.34 &-3.20 \\
LO Peg &K3 V & $\sim$25 & 9.25 &30&70 & 0.41    & 29.70 & -3.11 \\
\hline
\end{tabular}

\end{center}

\end{table*}

\emph{XMM-Newton} observations of YY~Gem are described by \cite{stelzer_2002}, and those of BO~Mic by \cite{wolter_2008}, and those of AB~Dor by \cite{lalitha_2013}, while
there appear to be no publications of the \emph{XMM-Newton} observations of PZ~Tel and HK~Aqr.
To avoid any biases in the estimated coronal properties we carry out a detailed analysis of all these stars similar to our analysis of LO~Peg and obtain their coronal properties such as emission measures, abundances, and coronal temperatures using our global fitting approach.

We simultaneously fit the medium-resolution EPIC data and high-resolution RGS data of each of the ultra-fast rotators with a combination of APEC models, using three temperature components, we allow the temperature, emission measure and abundances to vary as a free parameters. To obtain the plasma properties we use the total emission measure, EM~=~$\sum_i$EM$_i$ i.e. the sum the emission measure of each temperature component and 
the total temperature T =$\sum_i$  $\frac{T_i\times EM_i}{EM}$, i.e., the 
emission-measure weighted sum of each temperature component. 

The thus obtained coronal properties of our sample of ultra-fast rotators are listed in 
Tab.~\ref{coroprop}, where we also provide the (relative) iron abundance ($\frac{Fe}{Fe_{\odot}}$), 
ratio of neon-to-iron ($\frac{Ne}{Fe}$) abundance relative to solar photosphere, and the
ratio of neon-to-oxygen abundance ($\frac{Ne}{O}$).  Our analysis suggests that, all the sample stars consistently show an inverse FIP effect. 
We also note that the abundance ratios $\frac{Ne}{Fe}$ relative to solar-photosphere for LO~Peg, HK~Aqr, AB~Dor, BO~Mic and YY~Gem are $\sim$5, however, for PZ~Tel this ratio is only $\sim$~2. 

 \begin{table*}[!h]
\begin{center}
\caption{\label{coroprop} Observation ID and coronal properties of our sample of ultra-fast rotating low mass stars.}
\begin{tabular}[htbp]{lcccccccccccc}
\hline
\hline
Star & Obs. ID&T    &log EM 	& $\frac{Fe}{Fe_{\odot}}$ & $\frac{Ne}{Fe}$& $\frac{Ne}{O}$\\
        & &MK &  $10^{53}$cm$^{-3}$  & 		                                            &                             &          \\ 
\hline
AB Dor &0602240201&  11.49$^{+0.49}_{-0.54}$ & 10.18$^{+0.98}_{0.98}$ &   0.23$\pm$0.02&5.01$\pm$0.81& 0.44$\pm$0.06\\
BO Mic &0400460301& 11.78$^{+0.72}_{-0.64}$  & 8.11$^{+1.32}_{-1.25}$ &   0.21$\pm$0.07&5.69$\pm$1.14& 0.39$\pm$0.06 \\
HK Aqr &0202360101&  8.42$^{+0.67}_{-0.70}$ &0.93$^{+0.11}_{-0.10}$  &   0.17$\pm$0.08&5.47$\pm$0.73 &  0.36$\pm$0.04     \\
YY Gem & 0123710101& 10.20$^{+0.31}_{-0.29}$ & 3.38$^{+0.15}_{-0.14}$ &  0.29$\pm$0.05&5.35$\pm$0.67& 0.33 $\pm$0.05\\ 
PZ Tel&    0203060201&11.15$^{+0.37}_{_0.35}$&16.32$^{+1.42}_{-1.38}$ &  0.27$\pm$0.03 &2.03$\pm$0.41& 0.76$\pm$ 0.13\\
LO Peg & This work & 9.98$^{+0.47}_{-0.52}$ & 4.51$^{+0.44}_{-0.49}$&  0.23$\pm$0.02  &5.85$\pm$0.70& 0.54$\pm$0.07\\
\hline
\end{tabular}

\end{center}
\vspace{0.02cm}
\footnotesize{Note: For consistency we carried out a detailed analysis of each of the target and obtained the X-ray coronal properties using similar APEC models. \textbf{The abundance ratios $\frac{Ne}{Fe}$  are relative to the solar photosphere, while we obtain the $\frac{Ne}{O}$ ratio using the method described by \citealt{drake_2005}, i.e., the ratio is coronal.} }\\

\end{table*}

\begin{figure}[!h]
\begin{center}
\includegraphics[width=9cm]{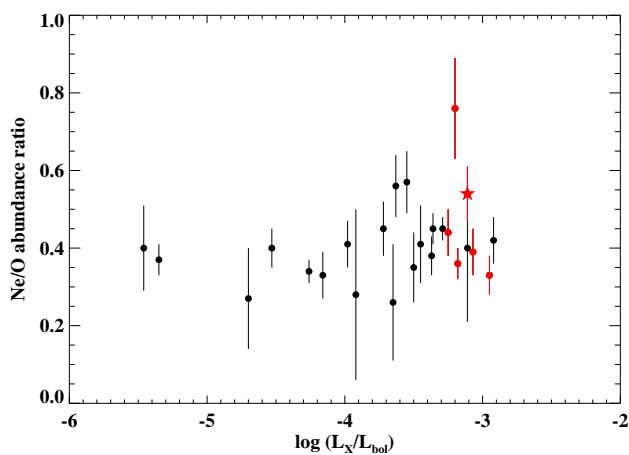}
\caption{\label{fig:neo}  Coronal Ne/O ratio for the sample of ultrafast rotators (red) with results from \cite{drake_2005} for  inactive to moderately-active solar-like stars (black filled circles). The red star symbol represents the LOPeg observation. }
\end{center}
\end{figure}

Ever since the controversy arose due to a disagreement between the helioseismology and the downward revision of solar abundance, the true neon abundance of the Sun and other stars has been debated. \cite{drake_2005} study the Ne/O ratios for a sample of 
low-to moderately-active stars and find the Ne/O value clustered 
around 2.5~-~4~times the solar Ne/O abundance ratio which is $\sim$0.17$\pm$0.05 \citep{young_2005, schmelz_2005}.
We derive the corresponding abundance ratios for our sample stars and compare them 
with the values obtained by \cite{drake_2005} and the classical solar value. 
In Fig.~\ref{fig:neo}, we plot the Ne/O ratios for our sample stars along with those stars 
previously reported by \cite{drake_2005} and find no particular trend for the fast rotators. 
Although our sample of highly active stars shows inverse-FIP effect biased coronae, 
we note that the Ne/O ratios for our sample stars are 0.43~$\pm$~0.15, similar to the value Ne/O =
0.40$\pm$0.08 for the stars studied by \cite{drake_2005}.   
The apparent dispersion in the mean abundance ratio for ultra-fast rotators is 
due to the outlier, PZ~Tel, which also has the largest emission measure of our
sample stars. 

\subsection{Flare loop properties}
\label{loopEMT}

Observations of solar and stellar flares have 
shown a correlation between the emission measure (EM) and 
the peak temperature (T) of flares, in the sense that both
change in a characteristic fashion during the course of
a flare.  Thus the EM-T diagram has become a useful diagnostic to 
estimate physical quantities that are not directly observable. 
Due to the lack of spatial resolution we assume that the observed flare  
occurs in a localised coronal region in a simplified geometry (single loop structure) 
and remains unchanged during the flare evolution.
The X-ray light curves during a flare are typically characterised
by a fast rise phase followed by a slower decay \citep{haisch_1983}.
The decay usually starts, when the heating decreases significantly.
As a result of the decay,  the plasma cools due to radiation and 
thermal conduction down to the chromosphere with some characteristic cooling time scale. 
The plasma cooling time explicitly depends on the confining 
loop structure, in particular, on the loop length and implicitly on the density \citep{reale_2002}. 
The decay time of the flare X-ray emission occurring inside a closed coronal structure scales with the plasma cooling 
time, which in turn scales with the length of structure confining the plasma. 
In other words, the longer the decay, the larger is the structure.
To summarise, the loop thermodynamic decay time scale, derived by \cite{serio_1991},
is given by 

\[
\tau_{s} = \alpha \frac{L}{\sqrt{T_{0}}} ~ = 120 \frac{L_{9}}{\sqrt{T_{0,7}}},
\]
where $\alpha=3.7 \times 10^{-4}$ cm$^{-1}$s\,K$^{1/2}$, $T_{0} ~(T_{0,7})$ is the loop maximum temperature in units of $10^7$ K 
and $L~(L_9)$ the loop half-length in units of $10^9$ cm. 
For stellar flares the loop length cannot be measured, however,
from theoretical models of a flaring loop a EM-T diagram
can be computed.  The important quantity is the 
slope $\zeta$, measured from the observable flare trajectory 
in the EM-T diagram. \cite{reale_2007} compute the loop length $L$ with the
formula 
\[
 L  = \frac{\tau_{LC} \sqrt{T_{0}}}{\alpha F(\zeta)} ~~ or ~~ L_{9}= \frac{\tau_{LC}
 \sqrt{T_{0,7}}}{120 F(\zeta)} ~~~~~  \zeta_{min}<\zeta\leq\zeta_{max} ~ ,
\]
where $\tau_{LC}$ is the decay time derived from the light curve, 
and the unit-less correction factor is $F(\zeta)$ is defined through
$F(\zeta) = \frac{c_{a}}{\zeta - 
\zeta_{a}} + q_{a}$.  According to \cite{reale_2007}, the coefficients  $c_{a}$, $\zeta_{a}$, 
and $q_{a}$ depend on the energy response of the 
instrument, and the same authors estimate the values for XMM/EPIC to be $c_{a} = 0.51$, $\zeta_{a} = 0.35$, and $q_{a} = 1.35$.
 
 \begin{figure*}
\begin{center}
\includegraphics[width=9cm]{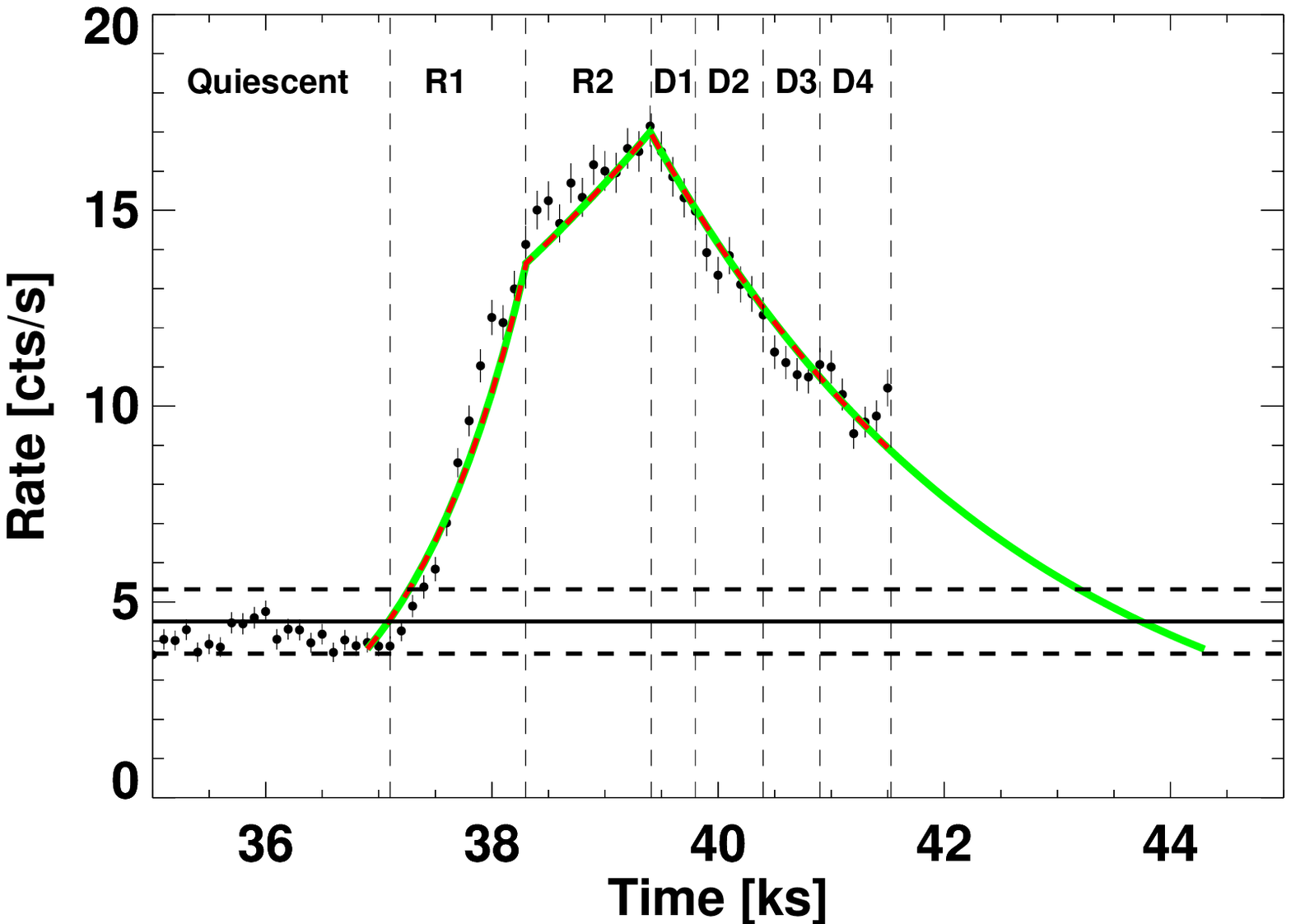}
\includegraphics[width=9cm]{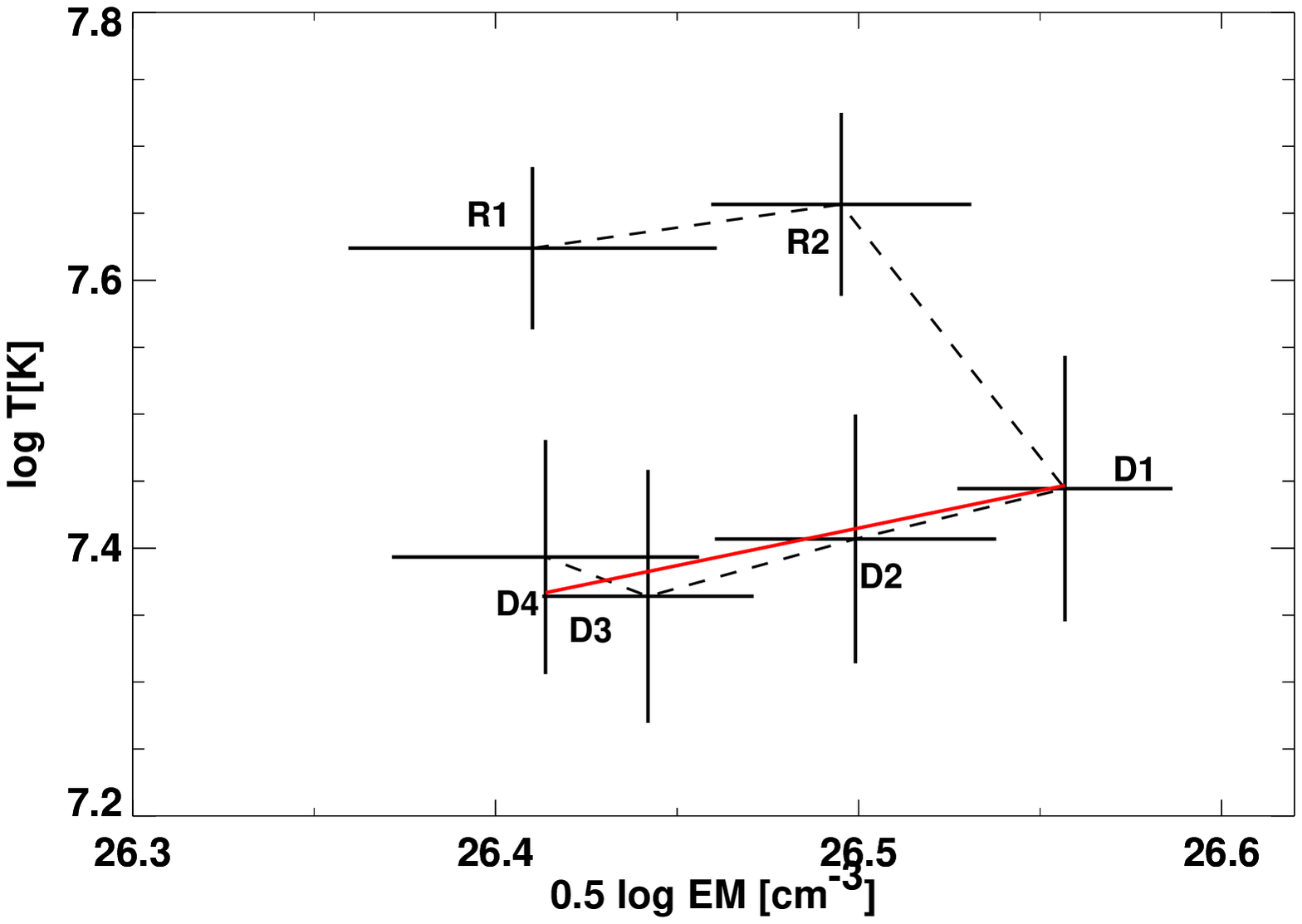}

\caption{\label{fig:mod} Upper panel:  The EPIC LO~Peg light curve (data points) overlaid with a simple exponential flare model (green curve). The vertical dashed lines mark pre-flare quiescence, rise (~R~) and decay phase time bins (~D~). Lower panel: Flare evolution in the EM-T diagram with corresponding time bins marked for reference.  } 
\end{center}
\end{figure*}

Using the following steps we then estimate the flare loop lengths:
\begin{itemize}
\item $F(\zeta)$ (slope of the EM-T evolution) : 
We divide the flare data into several time bins and 
obtain the emission-measure and temperature for the individual bins. 
In Fig.\ref{fig:mod} (upper panel), we show the EPIC pn light curve of the large flare observed on LO~Peg together with the time intervals used for our spectroscopic analysis; 
 the quiescent phase is also marked for reference.
The first two spectra cover the flare rise and the following time intervals cover the different phases in the flare.
Each of these spectra are fitted with a 4-T APEC model including the quiescent emission, i.e., the first two temperature components are fixed 
to the plasma properties of the quiescent phase before the flare. 
Thus we allow the third and fourth temperature component to vary independently, 
which accounts for the overall X-ray emission during the flare.
In Fig.~\ref{fig:mod} (lower panel), we show the evolution of the fit parameters
EM and T and measure a slope $\zeta$ for the decaying phase of 
$\zeta$ =0.56$\pm$0.35 in the EM-T plane. 
 
\item T$_0/T_{0,7}$ (Flare peak temperature):  \textbf{The intrinsic flare peak temperature is 
obtained by applying $T_{0} (T_{0,7}) = \xi T_{obs}^{\eta}$ (in units of $10^{7}$ K) to the 
observed maximum temperature; here the coefficients $\xi$=0.16 and $\eta$=1.16 have been
derived using the energy response of the \emph{XMM-Newton} EPIC detectors \citep{reale_2007}. 
Thus, the intrinsic flare temperature is estimated as 
$T_0 (T_{0,7})\approx$ 42.19 $\pm$ 5.86 MK. }

\item $\tau_{LC}$ (Flare decay time): 
We obtain the decay time by modeling the flare light curve. We assume an exponential growth and decay given by 
$CR=CR_{flare} \times e^{\frac{t-t_{flare}}{\tau_{rise/decay}}}$
with the observed e-folding time of the flare's light curve determined by fitting the light curve from the peak of the flare, until the count rate has reached
10$\%$ of the peak level. 
We extrapolate the light curve during the decay phase assuming an
exponential decay of the count rate, since our LO~Peg observations 
do not cover the entire flare, and 
obtain a decay time $\tau_{LC} \sim$3.26$\pm$0.43 ks. 
\end{itemize}

We  estimate a loop half length L $\sim$ 2.06$\pm$1.38 $\times10^{10}$ cm;
assuming a stellar radius of 0.72~$\pm$~0.10~$R_{\odot}$, 
we find a loop length L~$\sim$~0.41R$_{\star}$. 
Comparing the estimated loop length with the pressure scale height $H_p$, 
defined as $H_p$ = $\frac{2kT}{\mu g}$, where T is the plasma temperature 
in the loop,  $\mu$ is the molecular weight and g is the surface gravity of LO~Peg  
(log g$\sim$4.5$\pm$0.5, \citealt{pandey_2005}), and substituting 
these values in the expression for $H_p$, we find a value of 
$\sim 3.6 \times 10^{11}$ cm; thus the loop length obtained for LO~Peg flare 
is smaller than the pressure scale height of the flaring plasma. 

\subsubsection{Flare energetics}

A detailed assessment of the energy radiated in X-ray can be obtained using a simple integration of the instantaneous X-ray luminosity over the flare duration ($\sim$ 5 ks),
yielding a total energy $E$ radiated in the X-ray band of E $\sim$7.3$\times10^{33}$ erg. 
Observations of solar flares show that the density of the flaring loop during the peak and the decay phase of the flare are close to the quasi-steady state values; we now assume that this is also the case for the flare on LO~Peg.
One of the simplest model to understand the stellar flare energetics is to assume that the heating function is constant in space, i.e., assume that it is independent of the local values of the flare loop temperature and pressure.

In this case we can apply the scaling laws for static loops derived by \cite{rosner_1978}, i.e., the relation
$\frac{dH}{dV dt } \simeq 10^5 p^{7/6} L^{-5/6} $ holds, where
$L$ is the estimated loop length ($\sim$~2.06~$\pm$~1.38 $\times 10^{10}$ cm) and p$_0$ is the (constant)
pressure appearing in the other scaling-law, T$_{max}$= $1.4\times10^3$ (p$_0$ L)$^{\frac{1}{3}}$.  
The total heating rate during flare peak is then
$\frac{dH}{dt} \simeq \frac{dH}{dV dt} \times V \approx 9.4\times 10^{30} $erg/s, where we estimate the volume $V$ from the measured
emission measure and density.  
Thus, the obtained heating rate is a factor of $\approx$ 6 higher than the flare's X-ray luminosity 
obtained from spectral fitting.  This difference in the heating rate and the flare X-ray 
luminosity is compatible with the fact that during a flaring state, the X-ray emission is only one of 
the possible energy loss terms; at flare temperatures we expect the energy losses by thermal 
conduction to be quite high.


\section{Summary and conclusions}

With its X-ray luminosity of L$_X$=5.1$\times$10$^{29}$ erg/s in 0.2-10 keV band
the low-mass ultrafast rotator LO~Peg attains an activity level of 
log $\frac{L_X}{L_{bol}}$=-3.1, close to the saturation limit as expected,
yet its (quiescent) X-ray light curve does show variability at $\sim$30-40$\%$ level,
possibly due to rotational modulation.  A large flare is observed toward the end of 
our observation, which leads to significant spectral changes in the X-ray emission
as observed for other stars. 
At the onset of the flare, the optical emission peaks before the soft X-ray 
emission peak, and the recorded OM light curve follows approximately the time 
derivative of the soft X-ray emission, suggesting the presence of the Neupert effect and
and the evaporation of chromospheric material as cause for the soft X-ray flare.
We also find different maximum times for the harder and softer parts of the
observed X-ray emission, indicating the thermal cooling of the heated plasma. 
The XMM-Newton data thus indicate that the general features of solar flare models also
seem to apply to the flare observed on LO~Peg.

The overall X-ray properties of LO~Peg are quite similar to those of 
other active stars with dominant plasma components at temperatures of 
3, 7.5 and 20~MK during quiescence and 3.5, 12 and 32~MK during the flare,
when the emission measure of the hotter plasma increases significantly and 
higher temperatures are reached, while the cool plasma is only marginally changed.  
From the recorded X-ray light curves we infer loop lengths of the flaring plasma somewhat 
smaller than half the stellar radius, assuming that the flare has a simple geometry.

The coronal elemental abundances of LO~Peg show an inverse-FIP effect.  
These abundances were obtained relative to solar-photospheric abundance, since 
there exist no measured photospheric abundances of LO Peg.
A comparison of X-ray properties like the coronal temperature structure and the abundance pattern of LO~Peg with other active fast rotators shows them to be very similar. The determined coronal Ne/O abundance 
ratios of LO~Peg are also similar to other low-mass ultrafast rotators. We find no variation of 
the Ne/O ratio with activity level for our sample of very active rapid rotators. 

\begin{acknowledgements}
This work is based on observations obtained with XMM-Newton, an ESA science mission with instruments and contributions directly funded by the ESA Member States and the USA (NASA). S.L. acknowledges support from the DST INSPIRE Faculty fellowship. 
\end{acknowledgements}

\bibliographystyle{aa}
\bibliography{paper}

\end{document}